# Mist Generation Behavior in Ultrasonic Atomizer for Aerosol Jet® Printing

James Q. Feng [a], James D. Klett [b], and Michael J Renn [a]

[a] Optomec, Inc., 2575 University Avenue, #135, St. Paul, MN 55114, USA

[b] PAR Associates, Las Cruces, New Mexico, USA

james.q.feng@gmail.com

**Abstract**

Continuous ultrasonic atomization in a closed chamber is expected to generate a mist with an equilibrium droplet concentration and size distribution. Such a mist of microdroplets with controllable mist density has been used for Aerosol Jet® printing in the fabrication of a variety of additively manufactured microscale devices. Despite many unique capabilities demonstrated with the Aerosol Jet® printing technology, its ultrasonic atomization behavior appears to be rather sensitive to the ink properties with gaps in our understanding of the fundamental physics underlying its operation. In this work, we investigate some basic mechanisms in the Aerosol Jet® ultrasonic atomizer with a lumped-parameter kinetic coagulation model for highly concentrated mist. To mitigate the difficulty with unavailable knowledge about the complex turbulent flow inside the atomizer chamber, we present results for several orders of magnitude of the turbulent energy dissipation rates in order to examine a range of possibilities. The same approach is taken for analyzing the scavenging effect of the swirling bulk liquid. Our results also demonstrate the theoretical possibility for achieving a mist saturation condition where the mist output from the atomizer can become insensitive to process variables. As observed in experiments, such a saturated mist is highly desirable for Aerosol Jet® printing with maximized and well-controlled throughput in additive manufacturing.

## 1 Introduction

In Aerosol Jet® (AJ) printing as schematically shown in Fig. 1, functional ink materials are atomized into fine mist droplets and deposited to the substrate in the form of a high-speed mist stream with an impinging jet flow, based on the mechanism of inertial impaction of microdroplets with diameters typically ranging from 1 to 5 μm (Renn et al. 2002; Binder et al. 2014; Feng and Renn 2019). With a consistently supplied mist flow from the atomizer, the AJ deposition head is also equipped with a shuttering mechanism to switch the mist jet flow on and off to enable realistic feature printing (Feng 2023). It enables effective additive manufacturing of microscale devices for various industrial applications (cf. Zollmer et al. 2006; Hedges et al.

2007; Kahn 2007; Christenson et al. 2011; Paulsen et al. 2012; Renn et al. 2017; Wilkinson et al. 2019; Feng et al. 2019; Germann et al. 2023).  Any liquid material that can be atomized (or aerosolized) into a mist of microdroplets (e.g., with diameters ranging from 1 to 5 µm) can be printed with AJ.   The performance of the atomizer plays a key role in AJ operation.  One of the atomizers in the current AJ systems—called the "ultrasonic atomizer"—utilizes a megahertz ultrasonic transducer to excite short wavelength (on the order of microns) capillary waves that become destabilized to generate fine droplets.  Under appropriate conditions, the bulk liquid in the ultrasonic atomization chamber is so agitated by the ultrasonic energy that it swirls vigorously in the mist filled space, which also contributes to a scavenging mechanism for mist droplet removal, while creating more free surface area to enhance the capillary-wave based droplet generation (as illustrated in Fig. 1).

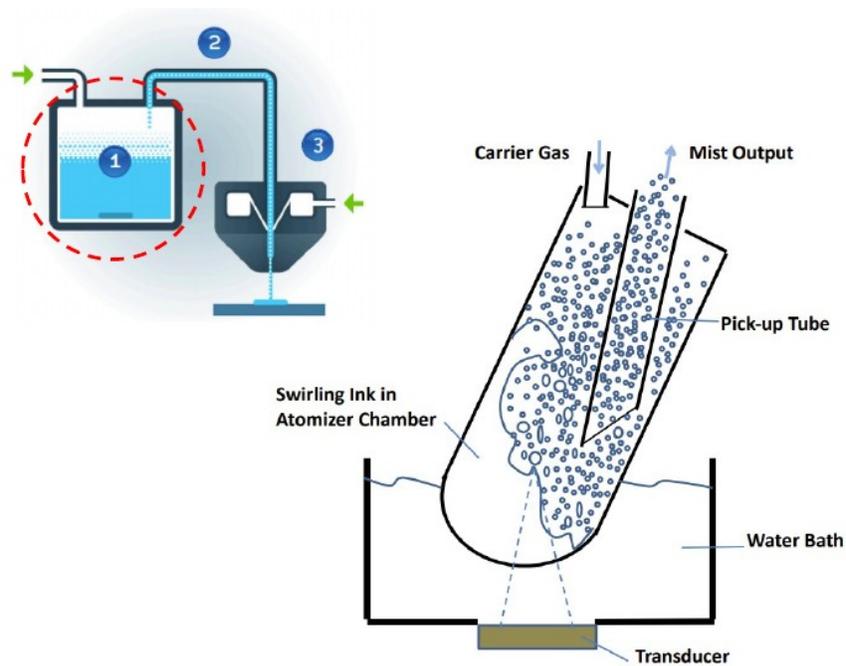

Fig. 1  Schematic of the Aerosol Jet® printing system: (1) Atomizer for generating mist of functional ink with constant mist density and droplet size distribution; (2) Mist transport channel; (3) Deposition head to form a high-speed collimated mist jet in an aerodynamic focusing nozzle with sheath gas for depositing ink on substrate (cf. Feng and Renn 2019), and diagram of an ultrasonic atomizer with adjustable transducer power (e.g., by varying the electric current to the transducer driver board at a fixed voltage.)

The ultrasonic atomizer in AJ systems usually works well with liquids of low viscosities (e.g., < 10 cp).  It can generate consistent mist output when the temperature of the atomization chamber as well as the ink therein is under adequate control, with simple mist transport and management for AJ ink deposition (cf. Feng and Renn 2019).  Similar devices (often called ultrasonic nebulizers) have also been widely used for drug delivery in aerosol inhalation therapy (Miller 2009). Due to its importance in various applications, the mechanism for ultrasonic atomization has become a subject of extensive research (Lang 1962; Fogler and Timmerhaus 1966; Topp 1973; Donnelly et al. 2005; Collins et al. 2012). In view of the commonly described ultrasonic droplet generation mechanism from destabilized capillary waves, it is theoretically expected to have difficulties for atomizing viscous liquids. Because the droplet size correlates with the excited capillary wavelength, to generate microdroplets requires megahertz frequency vibrations of the ultrasonic transducer, such that the wave breakup process can complete in less than a microsecond; otherwise the process could be reversed in the oscillation cycle of the waves. Intuitively, higher liquid viscosity would make quick breakup of capillary waves more difficult due to the viscous retardation effect on amplitude growth of unstable waves. However, aside from the dependence on viscosity, the behavior of generated mist has rarely been systematically characterized in terms of mist droplet concentration and size distribution, which are directly relevant properties affecting AJ printing throughput as well as process control.

According to established theory, droplets generated from unstable capillary waves should have more or less the same size, i.e., the mist droplets are nearly monodispersed. With a transducer of 1.6 MHz frequency, the diameter of ultrasonically generated droplets is estimated to be about 2.4 µm for typical AJ inks (Feng and Renn 2019). In realistic AJ applications, the microdroplets are continuously generated with an equilibrated constant mist density and size distribution inside the ultrasonic atomizer chamber, with a small portion of them continuously transported out by a carrier gas flow to the AJ deposition nozzle for printing while the majority remains inside the chamber, interacting with others, coalescing, settling back, etc. (cf. Fig. 1). The ideal situation is to have highly concentrated mist droplets fill up the atomizer chamber, reaching an equilibrium such that the mist droplet size distribution as well as the output mist density can become sustainable for a reasonably long time of consistent printing.

Droplets in a concentrated mist are expected to collide and coalesce, forming larger droplets which would be more readily to settle under gravity. Gravitational settling is one of the major droplet removal mechanisms in the ultrasonic atomizer chamber. The bulk liquid ink during ultrasonic atomization is also vigorously agitated by the ultrasonic energy to swirl inside the chamber, removing mist droplets by the mechanism of scavenging. With continuous generation of primary droplets of diameters around 2.4 µm and droplets removal mechanisms due to gravitational settling and swirling bulk liquid scavenging, an equilibrated mist with constant density and droplet size distribution can be generated. The experimental results of ink material mass output shown in Fig. 2 demonstrate the possibility to minimize sensitivities to the ultrasonic transducer power change and ink solid fraction variation for well-controlled AJ printing, and also provide a realistic reference for model results comparison.

It has been well recognized that the process of liquid atomization in general is still poorly understood in terms of fluid dynamics, despite its importance in various applications. With numerous droplets moving at various velocities, reflecting light, obscuring clear views of the details, experimental observations can be extremely difficult. Dense mist in the atomization chamber with multiple droplet collisions and interactions also challenges quantitative analysis. Although modern computational methods for direct numerical simulation could offer detailed results, it is only capable of simulating a small part of the mist while simulating two-phase flows with deforming free surfaces requires highly sophisticated numerical techniques. Moreover, the boundary conditions for simulating the ultrasonic atomization process shown in Fig. 1 cannot be easy to define with reasonable confidence, even when the resources is available for performing a very costly direct numerical simulation. Hence for now we take an approach of assembling a rather simplified model to provide a practically useful initial description about the complex ultrasonic atomization process.

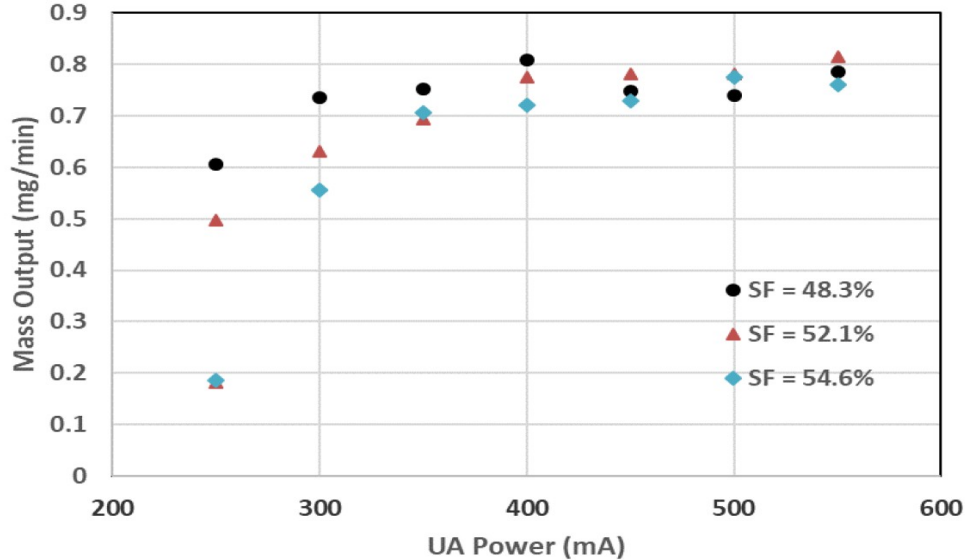

Fig. 2 Experimental results of solidified ink mass output of an aqueous silver nanoparticle ink by Novacentrix, for a mist flow rate 40 sccm, generated with an ultrasonic atomizer (UA) in the Aerosol Jet® Sprint system versus the electric current from a 48V power supply to the transducer driver board, demonstrating that above 350mA the mist output becomes insensitivity to both the UA power change and ink solid fraction (SF) variations.

In this work, we present a lumped-parameter model for simulating the mist density evolution in the AJ ultrasonic atomizer chamber for various values of the transducer power and parameters describing different droplet coagulation as well as removal mechanisms. Our aim in this study is to acquire an improved understanding of mist generation behavior and to provide some general guidance for atomization parameter adjustment to achieve the most desirable atomizer output.

**2 Model Description**

We assume a well-mixed mist inside the closed atomizer chamber that does not have a spatial variation, for simplicity. This model for the closed atomizer chamber implies that the rate at which the mist is taken out by the carrier gas flow to the deposition head (as illustrated in Fig. 1) is negligible in comparison to the mist generation rate during AJ printing. Moreover, the incompressible mist droplets are assumed to have volumes of $v_i = v_1 * i$ where "$v_1$" is the volume of the smallest droplets and "$i$" is an integer, implying that the effects of finer droplet generation, e.g., due to droplet breakup and splashing, are ignored here. Each increment of the integer $i$ indicates an increase of droplet volume by a unit of $v_1$ which usually corresponds to a relatively small increase of the droplet diameter. For example, increasing from $i = 1$ to 2, the droplet diameter would increase by a factor of $2^{1/3} = 1.26$; but from $i = 2$ to 3 and 3 to 4, the droplet diameters would increase by factors of $(3/2)^{1/3} = 1.14$ and $(4/3)^{1/3} = 1.10$, respectively. Corresponding to $i = 1, 2, 3, 8,$ and 27 we have droplets of diameter $d = 2, 2.52, 2.88, 4,$ and 6 µm. For this situation the droplet population evolves according to the discrete form of the source-augmented kinetic coagulation equation for droplets of volume $v_i$ (e.g., Klett, 1975) as

$$\frac{d n_i}{dt} = \dot{n}_i - S(i) n_i + \frac{1}{2} \sum_{j<i} K(i-j, j) n_{i-j} n_j - n_i \sum_{j=1}^{N} K(i, j) n_j, \quad (1)$$

where $n_i(t)$ denotes the number concentration of droplets of volume $v_i$ ; and $\dot{n}_i$ its primary generation rate, while $S(i)$ is a coefficient representing droplet removal mechanisms other than droplet coagulation (such as gravitational settling and scavenging due to swirling bulk liquid), and $K(i, j)\, n_j$ is the coagulation (or collection) rate of $j$ droplets with an $i$ droplet. The upper summation limit $N$ in (1) is taken to be a number large enough so that the resulting truncation of the spectrum for larger sizes is inconsequential.

For gravitational settling and scavenging due to swirling bulk liquid, the coefficient $S(i)$ can be expressed approximately as

$$S(i) = \frac{2\rho g}{9\mu h} \left( \frac{3 v_1}{4\pi} \right)^{\frac{2}{3}} i^{\frac{2}{3}} + \beta , \quad (2)$$

where $\rho$ denotes the mass density of droplet liquid (~ 2 g/cc for a typical metal nanoparticle ink in AJ printing), $g$ the acceleration of gravity (= 980 cm s$^{-2}$), $\mu$ the dynamic viscosity of gas phase (~ $1.8 \times 10^{-4}$ g cm$^{-1}$ s$^{-1}$), and $h$ the length scale for droplet settling which may be taken as the distance from center to the bottom of the ultrasonic atomizer chamber (e.g., $h = 3$ cm). The first term in (2) is just the characteristic reciprocal time required for a droplet falling in a quiescent fluid in the Stokes regime of negligible fluid inertia to settle to the bottom of the atomizer chamber. The Stokes flow model is a good approximation for droplet radii in the range from 1 to 10 µm. For lack of better knowledge, the effect of swirling bulk liquid scavenging is described by a proportionality parameter $\beta$ in (2), which is considered here to be independent of droplet size but can be a function of the transducer power; it represents an additional removal mechanism of mist droplets.

The mathematical form of the coagulation kernel $K(i, j)$ in (1) for binary collisions can be difficult to obtain, especially for the case of the simultaneous action of turbulent flows, Brownian diffusion, and gravitational acceleration between droplets of volumes $v_i$ and $v_j$ (e.g., Geng et al. 2013). It is still an active research area despite decades of effort (Saffman and Turner 1956; Pruppacher and Klett 1978; Williams 1988; Kruis and Kusters 1997; Wang et al. 2000; Park et al. 2002; Riemer and Wexler 2005; Zaichik and Alipchenkov 2008). The available formulas are therefore of an approximate nature, and typically involve some form of a lumped-parameter treatment.

2.1 Effects of Turbulence

Turbulent eddies produce transient local shear and acceleration flows. The former can result in "turbulent shear coagulation" and latter "turbulent inertial coagulation" of aerosol droplets. According to Williams (1988), the turbulent shear coagulation kernel can be expressed as

$$K_{TS}(i,j) = 5.65 \left(\frac{3v_1}{4\pi}\right) \left(\frac{\varepsilon}{v}\right)^{1/2} \left(i^{1/3} + j^{1/3}\right)^3, \qquad (3)$$

and the turbulent inertial coagulation kernel is given by

$$K_{TI}(i,j) = 1.38 \frac{\rho}{\mu} \left(\frac{3v_1}{4\pi}\right)^{4/3} \left(\frac{\varepsilon^3}{v}\right)^{1/4} \left(i^{1/3} + j^{1/3}\right)^2 \left|i^{2/3} - j^{2/3}\right|, \qquad (4)$$

where $v$ denotes the kinematic viscosity of gas phase (~ 0.15 cm$^2$ s$^{-1}$) and $\varepsilon$ the turbulent energy dissipation rate per unit mass of gas. It should be noted that (3) and (4) have the same functional form as those obtained by Saffman and Turner (1956) but with slightly different numerical values of the coefficients. To obtain these expressions, a geometric collision cross section has been assumed, in the absence of a theory for collision efficiencies in turbulent flow (e.g., Pruppacher and Klett 1978).

In the present work, we assume the value of $v_1 = (4\pi/3) \times 10^{-12}$ cc, corresponding to a droplet of 2 μm diameter. For $\varepsilon = 2000$ cm$^2$ s$^{-3}$, $\rho = 2$ g/cc and $\mu = 1.8 \times 10^{-4}$ g cm$^{-1}$ s$^{-1}$, $K_{TS}$ and $K_{TI}$ would become $7.53 \times 10^{-9}$ and $2.21 \times 10^{-9}$ cm$^3$ s$^{-1}$ for $i = 2$ and $j = 1$ (corresponding to droplet diameters 2.52 and 2.0 μm). If $\varepsilon$ is reduced to 200 cm$^2$ s$^{-3}$, we would have $K_{TS}$ and $K_{TI}$ becoming $2.38 \times 10^{-9}$ and $3.93 \times 10^{-10}$ cm$^3$ s$^{-1}$.

2.2 Effects of Gravity

Besides causing droplets to settle toward the chamber bottom, gravitational forces on droplets of different sizes also induce a relative velocity between them, which becomes an additional mechanism for collision and coalescence. The resulting gravitational coagulation kernel, with the concept of collision efficiency due to the hydrodynamic interaction between pairs of droplets being accounted for, may be expressed approximately by the form (Friedlander 1962; Klett 1975; Pruppacher and Klett 1978)

$$K_G(i,j) = \frac{\pi \rho g}{9\mu} \left(\frac{3v_1}{4\pi}\right)^{4/3} j^{2/3} \left(i^{2/3} - j^{2/3}\right), \qquad \text{for } i > j. \qquad (5)$$

This expression also assumes the interacting droplets are falling in the Stokes regime. As a reference, the value of $K_G$ would become $2.23 \times 10^{-10}$ cm$^3$ s$^{-1}$ for $i = 2$ and $j = 1$.

It is noteworthy that (4) and (5) do not contain any slip correction for non-continuum flow, which is justifiable in the present study for droplets of diameters greater than 2 μm because the Cunningham slip factor in this case would provide less than a 10% correction.

2.3 Effects of Brownian Motion

The Brownian coagulation kernel in the diffusion regime commonly used in the literature (Otto et al. 1999) takes the form

$$K_B(i,j) = \frac{2kT}{3\mu}\left(\frac{1}{i^{1/3}} + \frac{1}{j^{1/3}}\right)\left(i^{1/3} + j^{1/3}\right), \qquad (6)$$

where $k$ and $T$ are Boltzmann's constant (= $1.381 \times 10^{-16}$ dyne K$^{-1}$) and absolute temperature (K). As a reference, at $T = 300$ K the value of $K_B$ would become $6.22 \times 10^{-10}$ cm$^3$ s$^{-1}$ for $i = 2$ and $j = 1$.

2.4 Combined Coagulation Kernel

The combined effect of the simultaneous action of different coagulation processes is often assumed to be given by the simple summation of the respective processes acting in isolation. On the other hand, Saffman and Turner (1956) demonstrated that for the case of turbulence described by Gaussian statistics, a more accurate representation of combined effects is given by the root of the sum of squares of the separate coagulation rates. Later, Williams (1988) employed an approximate convective diffusion model to account for both Brownian and turbulent shear diffusion and obtained an interpolation formula for the combined effects. A more recent summary of best-available forms of combined coagulation kernels is provided by Geng et al. (2013). For the case of Brownian, turbulence and gravitational effects, the suggested overall kernel, which we adopt here, simply follows the ideas of Saffman and Turner (1956) and Williams (1988):

$$K(i,j) = \gamma_{BT}\left[K_B(i,j) + K_{TS}(i,j)\right] + \sqrt{K_{TI}(i,j)^2 + K_G(i,j)^2}, \qquad (7)$$

with $\gamma_{BT}$ being defined as (Williams, 1988)

$$\gamma_{BT} = \frac{1}{(1+3\chi^2)(1-0.5\pi\chi + \chi\tan^{-1}\chi)}, \quad \text{where} \quad \chi = \sqrt{\frac{0.90\pi\mu\sqrt{\varepsilon/\nu}}{kT}}\left(\frac{3v_1}{4\pi}\right)^{1/2}\left(i^{1/3} + j^{1/3}\right)^{3/2}.$$

For $\varepsilon = 2000$ cm$^2$ s$^{-3}$ and $\mu = 1.8 \times 10^{-4}$ g cm$^{-1}$ s$^{-1}$, $\gamma_{BT}$ becomes 1.016 for $i = 2$ and $j = 1$ with $v_1 = (4\pi/3) \times 10^{-12}$ cc. Even if $\varepsilon$ is reduced to 200 cm$^2$ s$^{-3}$, $\gamma_{BT}$ would still be 1.046. Thus, simply assuming a constant $\gamma_{BT} = 1$ is a justifiable simplification for our results with droplet diameters greater than 2 μm in the present study.

2.5 Governing Equation for Droplet Population Evolution

For convenience of numerical computation, it is desirable to measure the number concentration $n_i(t)$ in units of a nominal value $N_0$ such that all the parameters and variables would have values not too far from unity. In view of our empirical knowledge that a highly concentrated AJ mist could have a liquid volume fraction of about $5 \times 10^{-5}$ (Feng and Renn 2019), we estimate about $5.97 \times 10^6$ per cc droplets of 2.52 μm diameter. Thus assuming $N_0 = 10^6$ cm$^{-3}$ should be a reasonable choice. In equation (1), the net effect of normalizing $n_i$ by $N_0$ is to multiply the coagulation kernel $K(i, j)$ by $N_0$ to bring the value of $N_0 K(i, j)$ to around 0.001 s$^{-1}$ or so, while having $S(i) = 0.00807 \, i^{2/3} + \beta$ (in units of s$^{-1}$) according to (2). Hence, the actual value of number concentration for droplets of volume $v_i$ becomes the mathematical solution of dimensionless $n_i$ multiplied by $N_0$.

The generation rate $\dot{n}_i$ (in units of s$^{-1}$) for the mist droplet number concentration $n_i$ may be expressed as

$$\dot{n}_i = f(\alpha) c_i = \alpha c_i \, , \tag{8}$$

where the values of $c_i$ (in units of s$^{-1}$) define the relative primary generation rate of droplets of different sizes, while $\alpha \, (= p / p_0 - 1)$ is a dimensionless adjustable parameter describing the transducer power $p$ above a threshold value $p_0$ in the AJ ultrasonic atomizer. The value of $\alpha$ is assumed to be proportional to the mist generation intensity here as a first-order approximation.

Based on the observation of bulk liquid swirling intensity increasing with transducer power and usually correlating to mist generation intensity, it seems appropriate also to scale the parameter $\beta$ in (2) with $\alpha$, i.e. to have $\beta = B \alpha$ where $B$ (in units of s$^{-1}$) is another adjustable parameter used to define the scavenging rate by the swirling bulk liquid.

The governing equation for solving number concentration $n_i(t)$ (measured in units of $N_0$) is then written as

$$\frac{d n_i}{dt} = \alpha c_i - (0.00807 \, i^{2/3} + B\alpha) n_i + \frac{N_0}{2} \sum_{j<i} K(i-j, j) n_{i-j} n_j - N_0 n_i \sum_{j=1}^{N} K(i, j) n_j \quad , \tag{9}$$

where $\alpha$ and $B$ are freely adjustable parameters, with the coagulation kernel $K(i, j)$ given by (7). For lack of better knowledge, we assume $\varepsilon = 2000$ cm$^2$ s$^{-3}$ (which is not unreasonable in view of the values estimated for atmospheric turbulence by Saffman and Turner 1956) as the nominal value for the turbulent energy dissipation rate per unit mass of gas, unless otherwise specified.

## 3 Results and Discussion

In view of the theory for ultrasonic atomization (Lang 1962; Fogler and Timmerhaus 1966; Topp 1973; Donnelly et al. 2005; Collins et al. 2012), we assume a relatively narrow primary droplet size distribution in the source term (8). This amounts to having only a few nonzero values of $c_i$, e.g., $c_i = 0$ for $i > 4$. Although the computed solution to (9) is the number concentration $n_i(t)$ (in units of $N_0 = 10^6$ cm$^{-3}$), our results are mostly presented in practically relevant terms of volume fraction,

$$V(i) \equiv N_0 n_i \times v_1 \times i , \qquad (10)$$

with $v_1 = (4\pi/3) \times 10^{-12}$ cc, and volumetric mist density,

$$V_m \equiv \sum_{i=1}^{N} V(i) . \qquad (11)$$

In the present work, we compute terms with $i$ up to $N = 64$, beyond which adding more terms offers inconsequential effects to our results. The ordinary differential equation (9) is integrated simply by a forward Euler method, with the constant time step adjusted such that it is fine enough to ensure the numerically stable solution can reach a final equilibrium. To simulate the situation of suddenly switching on the transducer power at a given value of $\alpha$, the initial condition is set as $n_i(0) = 0$ (whereas the shutdown process is simulated with $\alpha = 0$ and $n_i(0)$ having an equilibrated set of values).

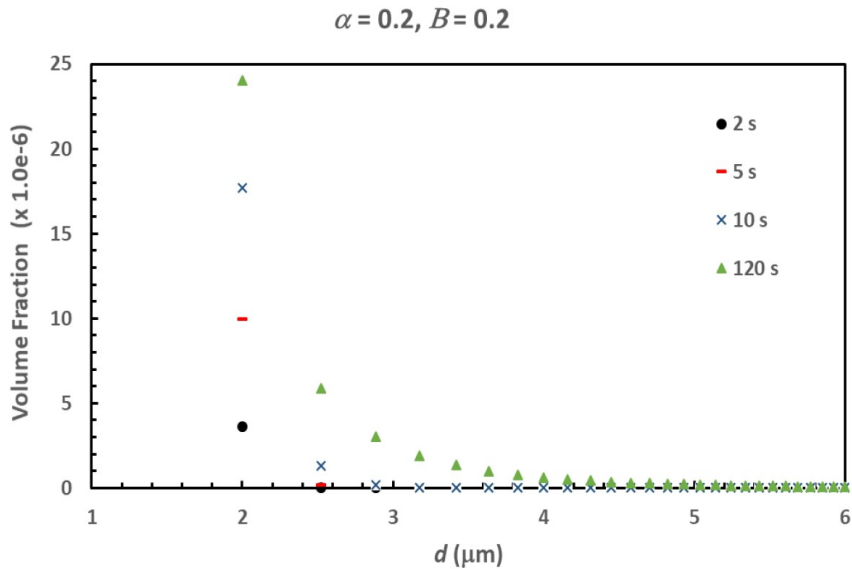

Fig. 3 Time evolution of volume fraction versus droplet diameter for $\alpha = 0.2$ and $B = 0.2$ s$^{-1}$, with $c_1 = 3$ s$^{-1}$ and $c_i = 0$ for $i > 1$ at $t = 2, 5, 10$, and 120 s. Here the droplet diameter is truncated at 6 μm in the plot only for enhancing visual clarity, not to mean 6 μm as the maximum droplet size in the computation.

3.1 Startup and Shutdown Behavior

If we have $c_1 = 3$ and $c_i = 0$ for $i > 1$, Fig. 3 shows the size distribution of droplets in the atomizer chamber, in terms of volume fraction versus droplet diameter $d = 2 \times i^{1/3}$ μm (for integer $i > 0$), when $\alpha = 0.2$ and $B = 0.2$ (s$^{-1}$). The volumetric mist density, calculated according to (11), becomes $V_m = 4.336 \times 10^{-5}$ in this case, the same as the observed high concentration mist in AJ printing output (e.g., about $5 \times 10^{-5}$, as alluded to by Feng and Renn 2019).

A mist generated from a monodispersed source at a rate of $6 \times 10^5$ droplets s$^{-1}$ per cc (or volume fraction about $2.52 \times 10^{-6}$ s$^{-1}$) will have its peak value of volume fraction at $d = 2$ μm increased to $V(1) = 3.63 \times 10^{-6}$, $9.97 \times 10^{-6}$, $1.77 \times 10^{-5}$, and $2.41 \times 10^{-5}$ at $t = 2, 5, 10$, and $120$ s, respectively. A broadened equilibrium distribution is attained at $t = 120$ s when the droplet generation rate mathematically equals the overall droplet removal rate. Clearly, larger droplets appear due to the coagulation mechanisms even with an idealized pure monodispersed droplet source.

In reality, a pure monodispersed droplet source does not exist; there is always a finite spread of droplet sizes no matter how narrow it may appear. For $c_1 = 1.0$, $c_2 = 0.8$, $c_3 = 0.3$, $c_4 = 0.1$ (s$^{-1}$), and $c_i = 0$ for $i > 4$, we obtain results in Fig. 4 for $\alpha = 0.2$ and $B = 0.2$ (s$^{-1}$) with $V_m = 4.818 \times 10^{-5}$ at equilibrium. The peak of volume fraction now appears around $d = 2.5$ μm, having values of $V(2) = 1.92 \times 10^{-6}$, $5.23 \times 10^{-6}$, and $9.04 \times 10^{-6}$ at $t = 2, 5$, and $10$ s, respectively.

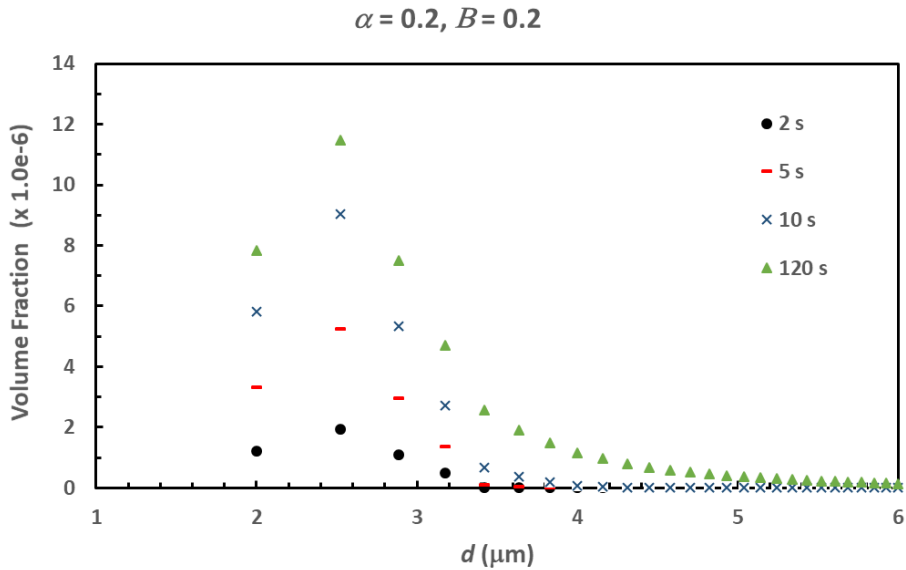

Fig. 4 As Fig. 3 but with $c_1 = 1.0$, $c_2 = 0.8$, $c_3 = 0.3$, $c_4 = 0.1$ (s$^{-1}$), and $c_i = 0$ for $i > 4$.

Majority of liquid volume is carried by droplets of $d$ from 2 μm to 3 μm, as typically observed with AJ printing (cf. Feng and Renn 2019). Again, droplets much larger than those generated primarily in the source (e.g., $d > 5$ μm) will appear in the equilibrated mist due to the inevitable coagulation process. The timescale for noticeable change in droplet size distribution due to coagulation seems to be more than 5 s. For convenience of comparison, all cases analyzed hereafter in this work will be computed with $c_1 = 1.0$ s$^{-1}$, $c_2 = 0.8$ s$^{-1}$, $c_3 = 0.3$ s$^{-1}$, $c_4 = 0.1$ s$^{-1}$, and $c_i = 0$ for $i > 4$.

To simulate the "shutdown" process, we can set an initial condition with equilibrated values of $n_i$ (e.g., those at $t = 120$ s for the case of Fig. 4) and $\alpha = 0$ for the forward Euler integration. Figure 5 shows a representative result for the shutdown process starting with the equilibrated mist of Fig. 4. Without an active source to generate fine droplets, larger droplets are continuously produced by coagulation at a rate higher than that of removal due to gravitational settling until the fine droplets are substantially depleted. It suggests that the shutdown process usually takes a few minutes for the volumetric mist density to reach a factor about 0.1 of the initial mist density, e.g., $V_m = 4.46 \times 10^{-6}$ and $1.75 \times 10^{-6}$ at 75 s and 120 s while $V(2) = 1.10 \times 10^{-5}$, $6.01 \times 10^{-6}$, $2.40 \times 10^{-6}$, and $3.90 \times 10^{-7}$ for $t = 1, 10, 30$, and 120 s, respectively in Fig. 5.

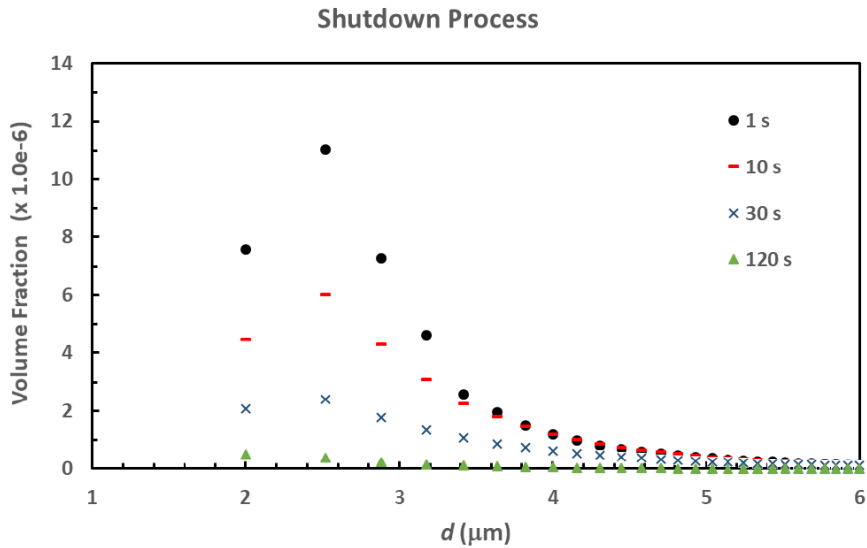

Fig. 5 Shutdown process with initial mist of $n_i$ from those at 120 s in Fig. 4 and $\alpha$ set to zero.

The transient evolution behaviors of volumetric mist density during startup and shutdown are illustrated in Fig. 6. In our numerical computation, we define the criterion for reaching the mathematical steady state as the root mean square value of $n_i$ variations between consecutive time steps becoming less than $10^{-7}$ for termination of the time integration. In reality, an equilibrium of mist density would be considered as established in a startup process when the AJ printing output variations become less than 10% or so. Thus, the practical timescale for mist to reach equilibrium density in startup could be much shorter than that indicated by the termination point for our time integration. According to Fig. 6, the actual timescale for mist density to reach the equilibrium value seems to be about 30 to 40 s, which is quite comparable to what has often been anecdotally observed in AJ printing during startup. For shutdown, Fig. 6 indicates that more than two minutes are required for the mist to settle out after the power is switched off, whereas the time integration is terminated at 240 s (i.e., 4 minutes), which again seems comparable to what is observed in AJ printing (e.g., usually around 3 to 4 minutes).

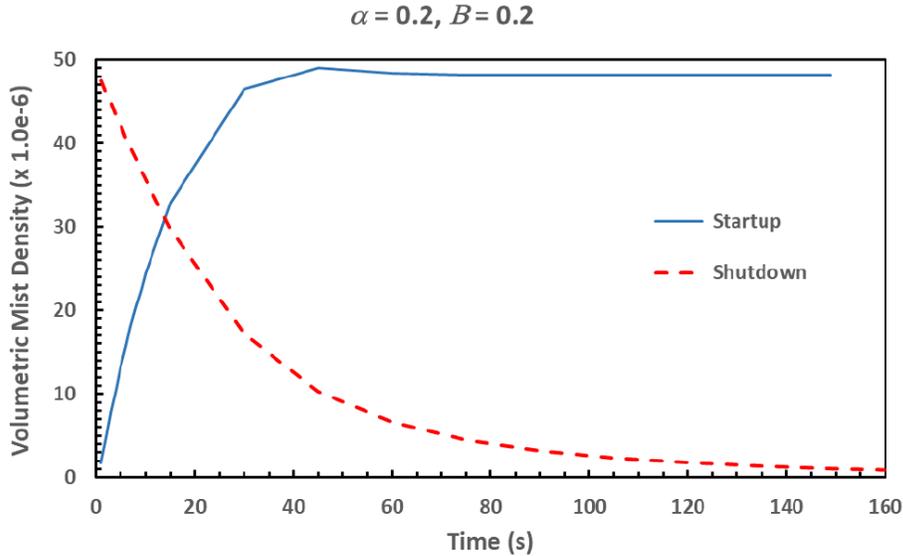

Fig. 6  Transient evolution of volumetric mist density for startup and shutdown with $\alpha = 0.2$ and $B = 0.2$ s$^{-1}$.

3.2 Evaluation of effects of $\alpha$ and $B$

Varying the values of $\alpha$ (the parameter describing the transducer power) and $B$ (the parameter for scavenging rate by the swirling bulk liquid) can affect the mist outcome. When the total volumetric mist density is kept about the same, Fig. 7 shows that relatively more of the larger droplets would appear in the equilibrated mist on reducing $\alpha$ and $B$. The case of $(\alpha, B) = (0.1, 0)$ indicates that Brownian coagulation and gravity effects can provide sufficient droplet removal mechanisms for establishing an equilibrium state, even without the swirling liquid.

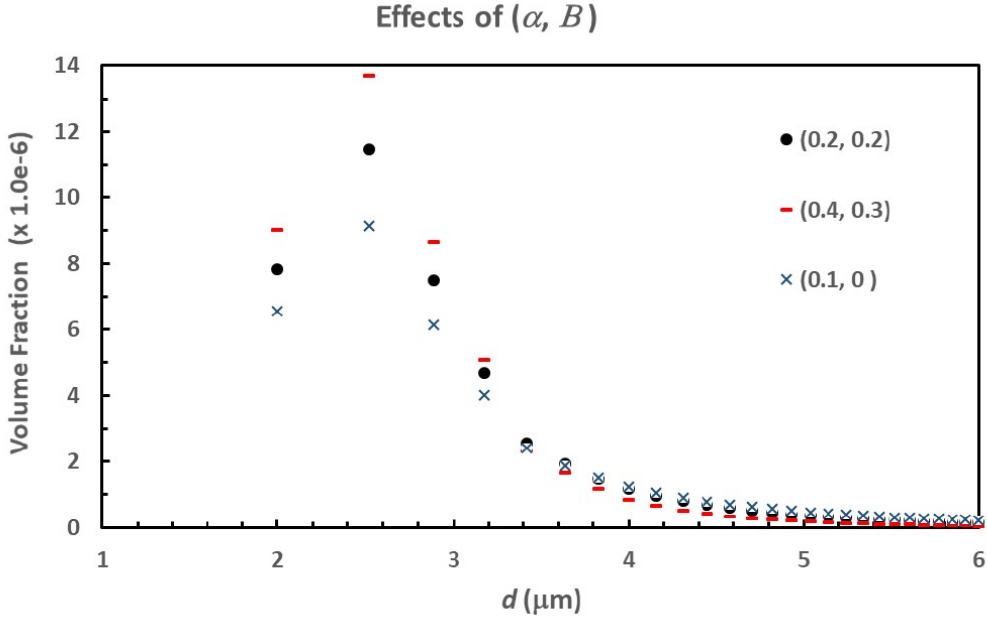

Fig. 7 Equilibrium droplet size distributions for ($\alpha$, $B$) = (0.2, 0.2), (0.4, 0.3), and (0.1, 0), with $B$ in units of s$^{-1}$, at $t$ = 149, 96, 211 s, respectively.

The numerical values given in Table 1 show more clearly the increased amount of larger droplets (i.e., $d$ = 6 μm with $V(27)$) as well as the longer mist equilibrating time upon reduction of $\alpha$ as well as $B$ for roughly the same volumetric mist density $V_m$. The effect of uniform scavenging (assumed proportional to the number concentration) by the swirling bulk liquid (represented by the value of $B$) appears to reduce the volume fraction of larger droplets more efficiently, because the loss of a larger droplet removes a much larger volume of liquid than a smaller droplet according to the cubic rule.

Table 1. Mist equilibrating time $t_e$, volumetric mist density $V_m$, peak volume fraction $V(2)$ for droplets of $d$ = 2.52 μm, and $V(27)$ for droplets of $d$ = 6 μm with various $\alpha$ and $B$.

| $\alpha$ | $B$ (s$^{-1}$) | $t_e$ (s) | $V_m$ (x10$^{-6}$) | $V(2)$ (x10$^{-6}$) | $V(27)$ (x10$^{-6}$) |
|---|---|---|---|---|---|
| 0.2 | 0.2 | 40 (149) | 48.18 | 11.47 | 0.15 |
| 0.4 | 0.3 | 30 (96) | 46.88 | 13.70 | 0.05 |
| 0.1 | 0 | 50 (211) | 45.73 | 9.16 | 0.22 |

It should be noted that the mist equilibrating time $t_e$ in Table 1 is given as an estimated practical timescale, with the termination point for time integration given in parentheses as a reference. Increasing $\alpha$ and $B$ tends to shorten the equilibrating time, but the amount of $t_e$ variation does not seem to be practically significant.

For AJ printing, the equilibrium volumetric mist density $V_m$ is one of the most important quantities, because it directly relates to the ink deposition throughput and therefore the productivity of the additive manufacturing equipment. In practice with AJ printing, it is also desirable to operate under the "mist saturation" condition with the value of $V_m$ becoming insensitive to all process variables including the transducer power. Theoretically, mist saturation could be possible when the mist density approaches its maximum value, beyond which the intensity of mist removal mechanisms (usually also increasing with mist density) becomes high enough to suppress any further increase of mist density. The phenomenon of mist saturation is fully controlled by the nature of coagulation and mist removal processes, and is apparently independent of the primary mist generation process (i.e., the value of $\alpha$ in our current model). Hence mist saturation is indicated by the "flat" portion of the $V_m$ versus $\alpha$ curve.

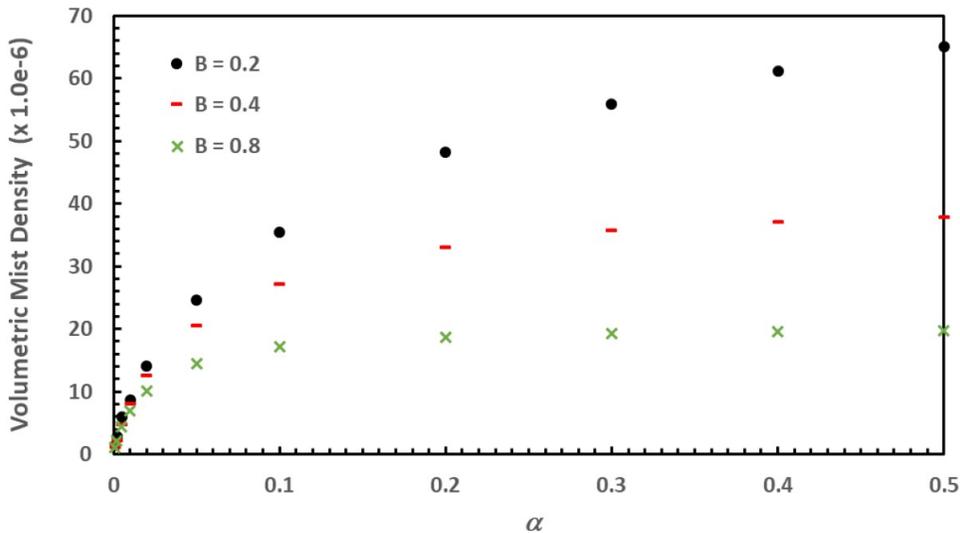

Fig. 8 Equilibrium volumetric mist density $V_m$ for various $\alpha$ at $B =$ 0.2 s$^{-1}$, 0.4 s$^{-1}$, and 0.8 s$^{-1}$.

Figure 8 exemplifies the possibility of achieving mist saturation by adjusting the values of $\alpha$ and $B$. To effectively flatten the $\alpha$ - $V_m$ curve, it is necessary to have an intensified swirling motion of bulk liquid (as represented by an increased value of $B$). Because the motion of swirling bulk liquid is directly driven by the transducer power, utilizing a more powerful transducer seems to be desirable. However, the reality can be much more complicated as it has been shown that for some liquids the mist output may become lower with increasing transducer power (Lozano et al. 2017). To model such situation would require a different form of $f(\alpha)$ in (8), such as a polynomial in $\alpha$, to relate the transducer power to an actual mist generation intensity. Such a case-by-case based analysis will not be pursued here, although it could be done. It is known that the atomization behavior in AJ operation is ink rheology dependent, with mist saturation observed for some well-formulated inks. The phenomenon of mist saturation never seemed to happen for inks with atomization difficulties and when the mist density is low, as consistent with the curves shown in Fig. 8. So, the importance of ink formulation should never be underestimated in optimizing AJ printing.

From the reference data presented in Fig. 2 with saturated solid silver mass output of 0.8 mg/min with 40 sccm mist flow rate for an ink of solid fraction about 0.5, the estimated value of $V_m$ would be about 2.2 x$10^{-5}$ as could be reasonably modeled by setting $B \sim 0.7$ s$^{-1}$ in view of the curves shown in Fig. 8.

3.3 Evaluation of Turbulent Intensity

Among several parameters in the coagulation model, the most uncertain one is the intensity of turbulence, i.e., the value of $\varepsilon$ (the turbulent energy dissipation rate per unit mass of gas) due to lack of detailed knowledge. Without much quantitative basis, we have chosen a hypothetical value of $\varepsilon = 2000$ cm$^2$ s$^{-3}$ for the turbulent energy dissipation rate per unit mass of gas, in our nominal case study. It can be useful at least for exploring the atomizer behavior in response to the external excitation $\alpha$ and associated parameter $B$ for mist scavenging effects due to the swirling bulk liquid. In principle, we may expect the turbulence intensity to also depend on the value of $\alpha$, because without an external excitation all the fluids in the atomizer chamber should remain in a quiescent state; but it is difficult to come up with a reasonable functional relationship at present.

Table 2. Mist equilibrating time $t_e$, volumetric mist density $V_m$, normalized peak volume fraction $V(2) / V_m$ for droplets of $d = 2.52$ μm, and $V(27) / V_m$ for droplets of $d = 6$ μm with various $\varepsilon$ at $\alpha = 0.1$ and $B = 0.2$ s$^{-1}$.

| $\varepsilon$ (cm$^2$ s$^{-3}$) | $t_e$ (s) | $V_m$ (x$10^{-6}$) | $V(2) / V_m$ | $V(27) / V_m$ |
|---|---|---|---|---|
| 2000 | 50 (195) | 35.54 | 0.237 | 0.0029 |
| 200 | 60 (240) | 43.84 | 0.299 | 0.00028 |
| 20000 | 30 (125) | 22.16 | 0.204 | 0.0052 |

In the absence of information on the actually realized values of turbulent intensity in the atomizer chamber, we can at least explore its approximate impact on our results by varying the value of $\varepsilon$ over a few orders of magnitude for otherwise the same conditions. For example, Table 2 and Fig. 9 show comparison among $\varepsilon = 2000$ cm$^2$ s$^{-3}$, 200 cm$^2$ s$^{-3}$, and 20000 cm$^2$ s$^{-3}$ for the case of $\alpha = 0.1$ and $B = 0.2$ (s$^{-1}$), in terms of $t_e$, $V_m$, $V(2) / V_m$ and $V(27) / V_m$ as well as equilibrated droplet size distribution. The values of $V(2) / V_m$ and $V(27) / V_m$ provide relative measures of the equilibrium spectra of droplets. As expected, more intensified turbulence tends to enhance the coagulation process for transferring liquid upward in volume in the droplet size spectrum. This leads to more efficient removal of mist liquid by gravitational settling and swirling bulk liquid, and therefore the volumetric mist density decreases with the magnitude of $\varepsilon$. The enhanced turbulent coagulation process can also speed up the mist equilibration process.

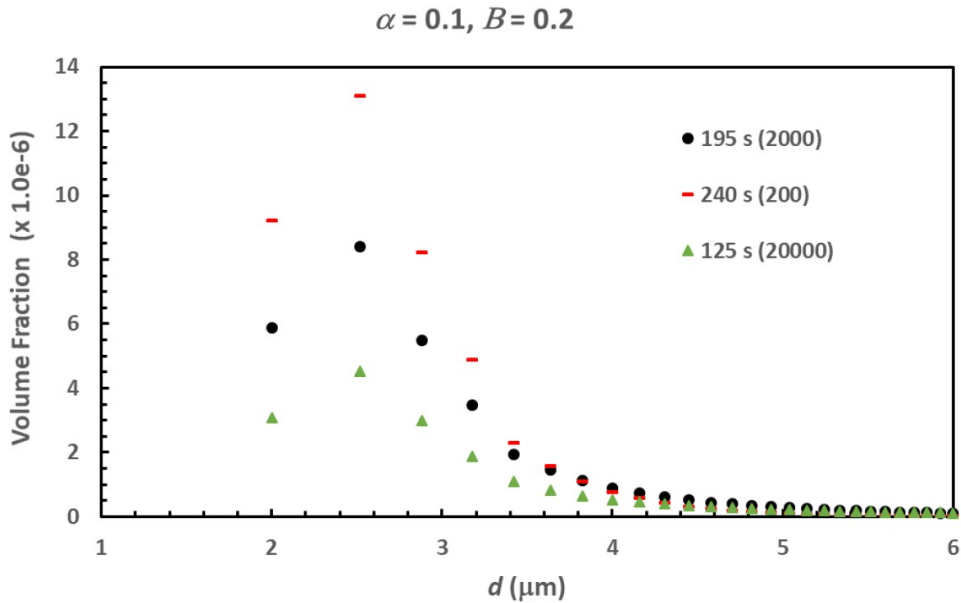

Fig. 9 As Fig. 6 but for $\alpha = 0.1$ and $B = 0.2$ s$^{-1}$ with $\varepsilon = 2000$ cm$^2$ s$^{-3}$ at $t = 195$ s, $\varepsilon = 200$ cm$^2$ s$^{-3}$ at $t = 240$ s, and $\varepsilon = 20,000$ cm$^2$ s$^{-3}$ at $t = 125$ s with the value of $\varepsilon$ given in parentheses

Even for $\varepsilon = 0$, the value of $V_m$, would still be 4.820 x10$^{-5}$ with $t_e \sim 75$ s for $\alpha = 0.1$ and $B = 0.2$ s$^{-1}$, quite comparable with those for $\varepsilon = 200$ cm$^2$ s$^{-3}$ in Table 2. Thus, within several orders of magnitude of $\varepsilon$ value variations at $(\alpha, B) = (0.1, 0.2)$, our modeling results seem to be generally reasonable in terms of the mist equilibrating timescale and equilibrium volumetric mist density (varying only by about a factor of 2).

**4 Concluding Remarks**

A model based on the discrete form of the source-augmented kinetic coagulation equation is formulated for simulating the ultrasonic atomization behavior in AJ printing. Despite the possibility of oversimplification of many intractably complicated fluid dynamic phenomena, the computed results of this lumped-parameter model appear to capture the essence of anecdotally observed atomization characteristics during AJ printing practice.  For example, much large droplets than the source spectrum can appear in equilibrated mist due to the nature of coagulation mechanisms, even for a monodispersed source.  By adjusting the values of the transducer power $\alpha$ and the parameter associated with swirling bulk liquid scavenging $B$, our model results show that a range of realistic values of mist density can be obtained with timescales for startup and shutdown matching what have often been observed in AJ printing.  Depending on ink properties (as represented by the values of $\alpha$ and $B$), it is also possible to obtain mist saturation when the mist density becomes insensitive to all process variables.  Our results show that an intensified swirling bulk liquid is necessary for achieving mist saturation with realistic mist density.  By varying the turbulent energy dissipation rate for over several orders of magnitude, our model results illustrate its relative insensitivity (about a factor of 2 changes in the mist density) to the unknown nature of turbulence in the atomizer chamber.  Although there is still plenty of room for fine-tuning the functional relationships among adjustable parameters, such efforts should await more experimental data, acquired with appropriate instrumental and measurement methods.  At the current stage, this lumped-parameter model can be useful for gaining a preliminary mechanistic understanding of the ultrasonic atomization behavior, even though it may not yet sophisticated enough to provide exact predictions.

**Statements and Declarations**

**Competing interests**  The authors have no competing interests to declare